\documentclass[manuscript]{aastex}

\shorttitle{Magnetic field restructuring associated with two successive solar eruptions}
\shortauthors{Rui Wang and Ying Liu}

\begin{document}

\title{Magnetic Field Restructuring Associated with Two Successive Solar Eruptions}

\author{Rui Wang\altaffilmark{1}, Ying D. Liu\altaffilmark{1}, Zhongwei Yang\altaffilmark{1} and Huidong Hu\altaffilmark{1}}


\altaffiltext{1}{State Key Laboratory of Space Weather, National Space Science Center, Chinese Academy of Sciences, Beijing, China; liuxying@spaceweather.ac.cn}


\begin{abstract}

We examine two successive flare eruptions (X5.4 and X1.3) on 2012 March 7 in the NOAA active region 11429 and investigate the magnetic field reconfiguration associated with the two eruptions. Using an advanced non-linear force-free field (NLFFF) extrapolation method based on the SDO/HMI vector magnetograms, we obtain a stepwise decrease in the magnetic free energy during the eruptions, which is roughly $20\%-30\%$ of the energy of the pre-flare phase. We also calculate the magnetic helicity, and suggest that the changes of the sign of the helicity injection rate might be associated with the eruptions. Through the investigation of the magnetic field evolution, we find that the appearance of the ``implosion" phenomenon has a strong relationship with the occurrence of the first X-class flare. Meanwhile, the magnetic field changes of the successive eruptions with implosion and without implosion were well observed.

\end{abstract}





\section{Introduction}

It is commonly believed that the coronal magnetic field plays a very important role during the eruptions
of solar flares and coronal mass ejections (CMEs). The magnetic free energy and helicity often change prominently during the process of these transient phenomena. Flares and CMEs derive their energy stored in the magnetic field, and in general the energy is released from an active region (AR). The energy released is just the magnetic free energy, which is the energy exceeding the potential field energy. A potential magnetic field structure is a minimal energy configuration of the magnetic fields. The causes for an AR non-potential configuration mainly include twisting and shearing of the magnetic field produced by the footpoint motion of the magnetic field lines on the surface of the photosphere, and flux emergence from underneath the photosphere. Sometimes even the newly emerged flux itself is non-potential. No matter what the cause is, it is a disturbance to the magnetic field and the magnetic energy and helicity may change accordingly.

During big flares the magnetic field often shows a rapid, irreversible change \citep[e.g.,][]{2012liuchang,2012wangshuo,2013petriea,2013wanghm}. The magnetic field becomes more horizontal after the eruption than before. This is explained by an ``implosion" theory \citep{2000hudson}. Specifically, during the eruption part of the magnetic field shrinks or collapses (``implodes") so that there is an overall decrease in the magnetic energy in the region of eruption. The contraction behavior of the magnetic field will happen when it loses the energy supporting its configuration, so as to achieve a new force balance.

The AR NOAA 11429 spawned a powerful X5.4 flare on 2012 March 7, which is the second largest flare eruption event since 2010. It is associated with a wide and fast CME of more than 2000 km s$^{-1}$ around 00:24 UT on March 7 \citep{2013liuxy,2014liuxy}. Of particular interest is that there were two successive eruptions, both of which are X-class flares. The second flare reached X1.3 less than 1 hr after the first X5.4 flare eruption. Related work has been done about AR 11429. Wang et al. \citeyearpar{2012wangshuo} and Petrie \citeyearpar{2013petrieb} both derive a stepwise increase in the magnetic field on the photosphere during the eruptions. Also, Wang et al. \citeyearpar{2012wangshuo} compute the Lorentz force and find that the stepwise decrease in the Lorentz force has a positive correlation with the peak soft X-ray flux, so they suggest that the CME mass can be estimated by the Lorentz force change.

In this paper, we use data from the Helioseismic and Magnetic Imager (HMI; \citealt{2012schou}) on board the Solar Dynamics Observatory (SDO) to investigate the magnetic free energy and magnetic helicity of this AR. In Section \ref{observations}, we show data analysis and use an advanced coronal magnetic field extrapolation method to understand the the coronal magnetic field topology. We also discuss the evolution of the magnetic free energy and magnetic helicity. In Section~3 we discuss the magnetic field reconfiguration during the successive eruptions. Section \ref{conclusion} summaries the conclusions.

\section{Observations and Data Analysis}\label{observations}

\begin{figure}
\epsscale{.80}
\plotone{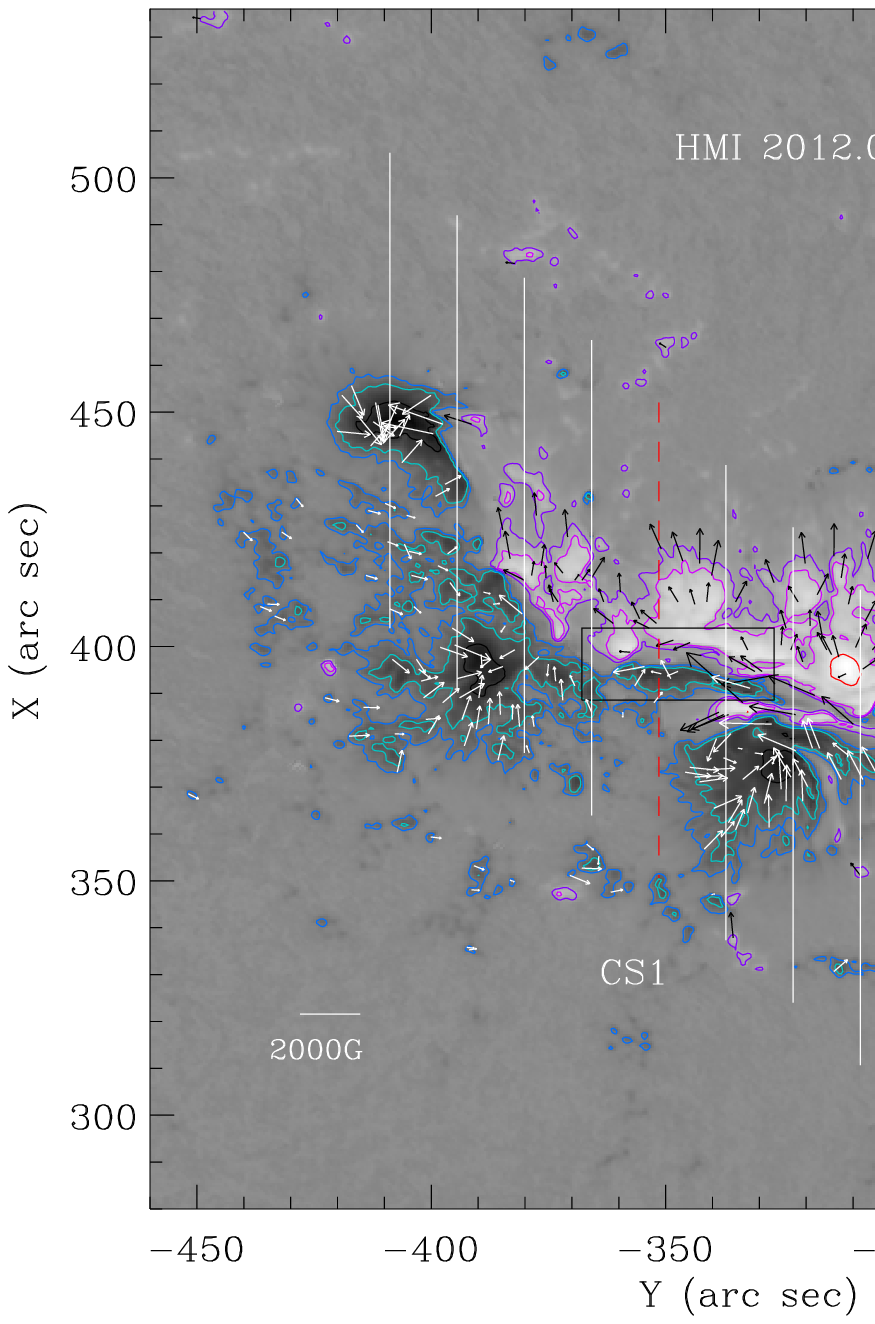}
\caption{Remapped HMI vector magnetogram for the region of AR 11429 as viewed from overhead. The vertical field ($B_z$) is plotted as the background. The black (white) arrows indicate the horizontal field ($B_h$) with positive (negative) vertical field components. The blue, green and black contours are plotted at -500, -1000, and -2000 G; the purple, pink, and red contours are plotted at 500, 1000 and 2000 G, respectively. The positions of ten uniform slices are marked as the white solid lines, and red (CS1) and blue (CS2) dashed lines. The black rectangle marks a region near the polarity inversion line, which is used in the subsequent analysis. The coordinate is in arc sec. The origin is relative to the center of the solar disk.\label{magmap}}
\end{figure}

The HMI instrument provides high time-resolution vector magnetic field data for NOAA AR 11429 with a 12 minute cadence and $0".5$ pixel size. In this investigation, we adopt the Lambert Cylindrical Equal Area (CEA) projected and remapped vector magnetic field data \citep{1990gary,2002calabretta,2006thompson}, as shown in Figure~\ref{magmap}, to investigate the evolution of the magnetic free energy and helicity of the two successive X-class flares on 2012 March 7. The X5.4 flare started at 00:02 UT, peaked at 00:24 UT, ended at 00:40 UT, and was companied by a CME of more than 2000 km s$^{-1}$; the X1.3 flare started at 01:05 UT, peaked at 01:14 UT, ended at 01:23 UT, and was accompanied by another CME of about 1800 km s$^{-1}$ \citep{2013liuxy}. The $180^\circ$ azimuthal ambiguity in the transverse field of the data is resolved by a minimum energy algorithm \citep{1994metcalf,2009leka}. We also adopt the HMI vector magnetic field data to calculate the magnetic helicity. In addition, we extract ten slices from the extrapolated coronal magnetic field data cube, which have the same area (72.5$\times$72.5 Mm), and their positions are shown in Figure~\ref{magmap}. With the ten slices we can compare the magnetic flux changes in different locations.

\subsection{Evolution of Photospheric Magnetic Field Near PIL}

\begin{figure}
\epsscale{.80}
\plotone{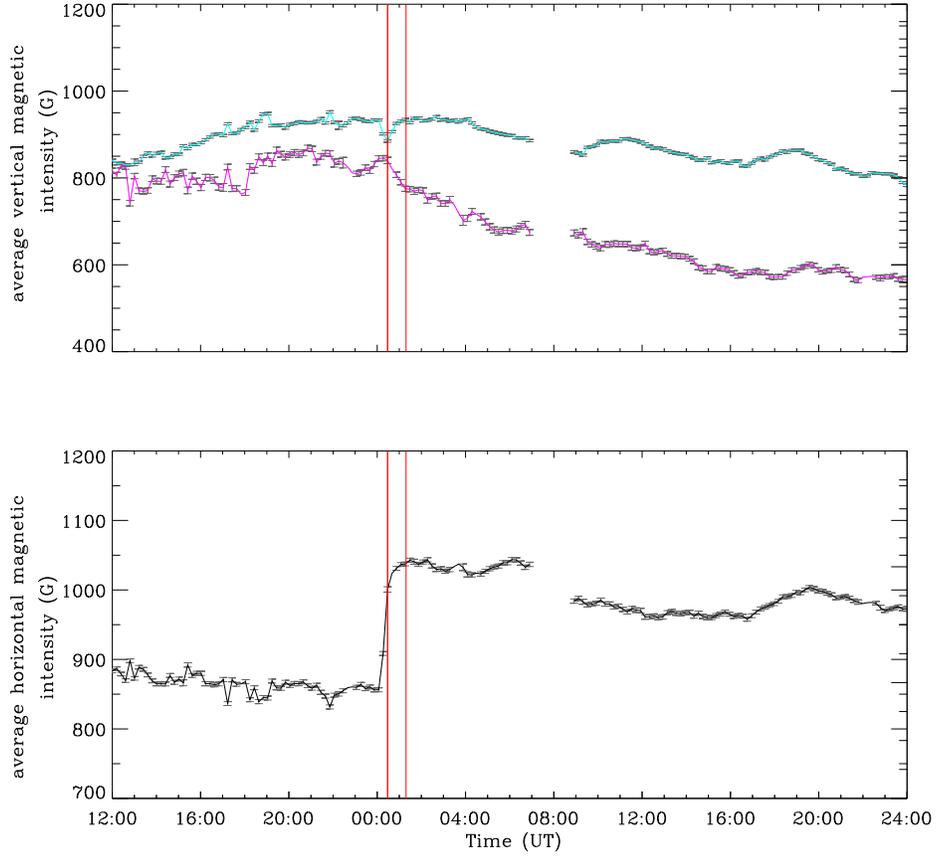}
\caption{Average vertical magnetic intensity (upper panel) and horizontal field (bottom panel) near the neutral line as a function of time. The cyan and purple lines represent positive and negative field components in the upper panel. The thicker and thinner vertical red lines represent the first and second GOES flare peak times, respectively. The uncertainties of the average magnetic field are plotted as error bars in $3\sigma$ level.\label{magfluxPIL}}
\end{figure}

We investigate the evolution of the photospheric magnetic fields near the polarity inversion line (PIL). Figure~2 shows the temporal profiles of the magnetic field changes. These plots of temporal changes are derived by calculating area integrals of the field components over the chosen photospheric areas along the PIL in 176 12-minute images, from 12:00 UT on March 6 to 23:48 UT on March 7, i.e.,
\begin{equation}
F_{PIL}=\int_{A_{PIL}}BdA,
\end{equation}
where $B$ is the magnetic field on the photosphere, and $A_{PIL}$ is the area marked by a black rectangle corresponding to the region near the PIL. The top panel of Figure~\ref{magfluxPIL} shows the average vertical magnetic intensity, and the cyan/purple lines represent positive/negative intensity, respectively. They have a similar trend. The average positive vertical intensity is a little higher than the negative one. During the eruptions, they seem to have an opposite change but not so obvious. Before 08:00 UT, there was a transient drop of the vertical and horizontal intensity. We think the data is fake. As we can see there was an interval of data missing before 08:00 UT. So the observation values are not real, just in a transition period, namely the value increases from zero to normal level. The error bars of the average magnetic intensity shown in Figure~2 are given by $3\sigma$ where $\sigma$ is the standard deviation of the HMI data in the region of the black rectangle. The bottom panel shows the horizontal fields. There was an obvious increase in the horizontal average magnetic intensity during the eruptions, and the most prominent part of the increase occurred during the first larger flare. Compared with the horizontal fields, the vertical fields had no such abrupt increase during the eruptions, and also the field strength is relatively weaker. Similar rapid enhancements of transverse fields during big flare eruptions are also found in other works \citep{1992Wanghm,2010wanghm,2012liuchang,2012wangshuo,2013petriea,2013wanghm}. According to \citet{2008ASPC..383..221H}, the reason that the magnetic field becomes more horizontal is that the coronal magnetic field contracts downward. The magnetic contraction could be explained by the ``implosion" theory of \citet{2000hudson}, and this will be discussed further in Section~\ref{discussion}.

\subsection{Evolution in Magnetic Free Energy}

Here, we obtain the coronal magnetic fields by adopting the Non-Linear Force-Free Field (NLFFF) method as proposed by Wheatland et al. \citeyearpar{2000wheatland} and extended by Wiegelmann \citeyearpar{2004wiegelmann} and Wiegelmann and Inhester \citeyearpar{2010wiegelmann}. We use the latest version of the NLFFF optimization code improved in 2011 \citep{2012wiegelmann} to extrapolate the coronal field from the observed vector magnetograms in a Cartesian domain. A preprocessing procedure has been used to remove most of the net force and torque from the data, so the boundary can be more consistent with the force-free assumption. Also, we obtain a potential field (PF) configuration from the same observation using the vertical component of the fields with the help of a Fourier representation based on Green's function \citep{1978seehafer}. The magnetic free energy ($E_{free}$) can be inferred by subtracting the PF energy ($E_{pot}$) from the NLFFF energy ($E_{nff}$),
\begin{equation}
E_{free}=\int_V \frac{B^2_{nff}}{8\pi}dV-\int_V\frac{B^2_{pot}}{8\pi}dV,
\end{equation}
where the energy is computed from the field strength within a certain volume $V$, and the subscripts $nff$ and $pot$ denote NLFFF and PF, respectively. The time resolution of the free energy time series is 1 hr, but there is a higher resolution of 12 min before and after the eruptions from 23:00 UT on March 6 to 02:00 UT on March 7.

\begin{figure}[H]\centering

\includegraphics[width=.32\textwidth]{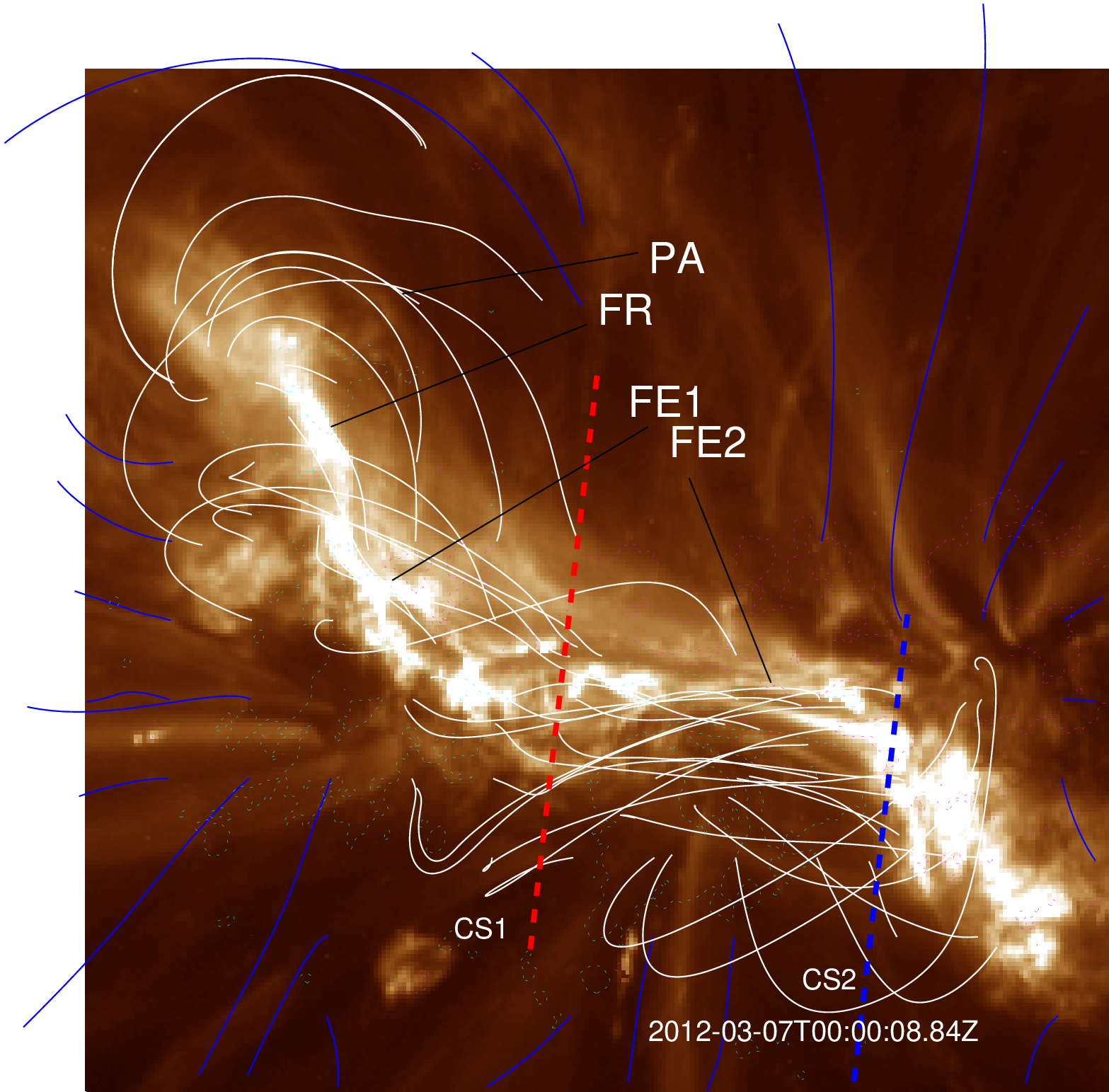}
\includegraphics[width=.32\textwidth]{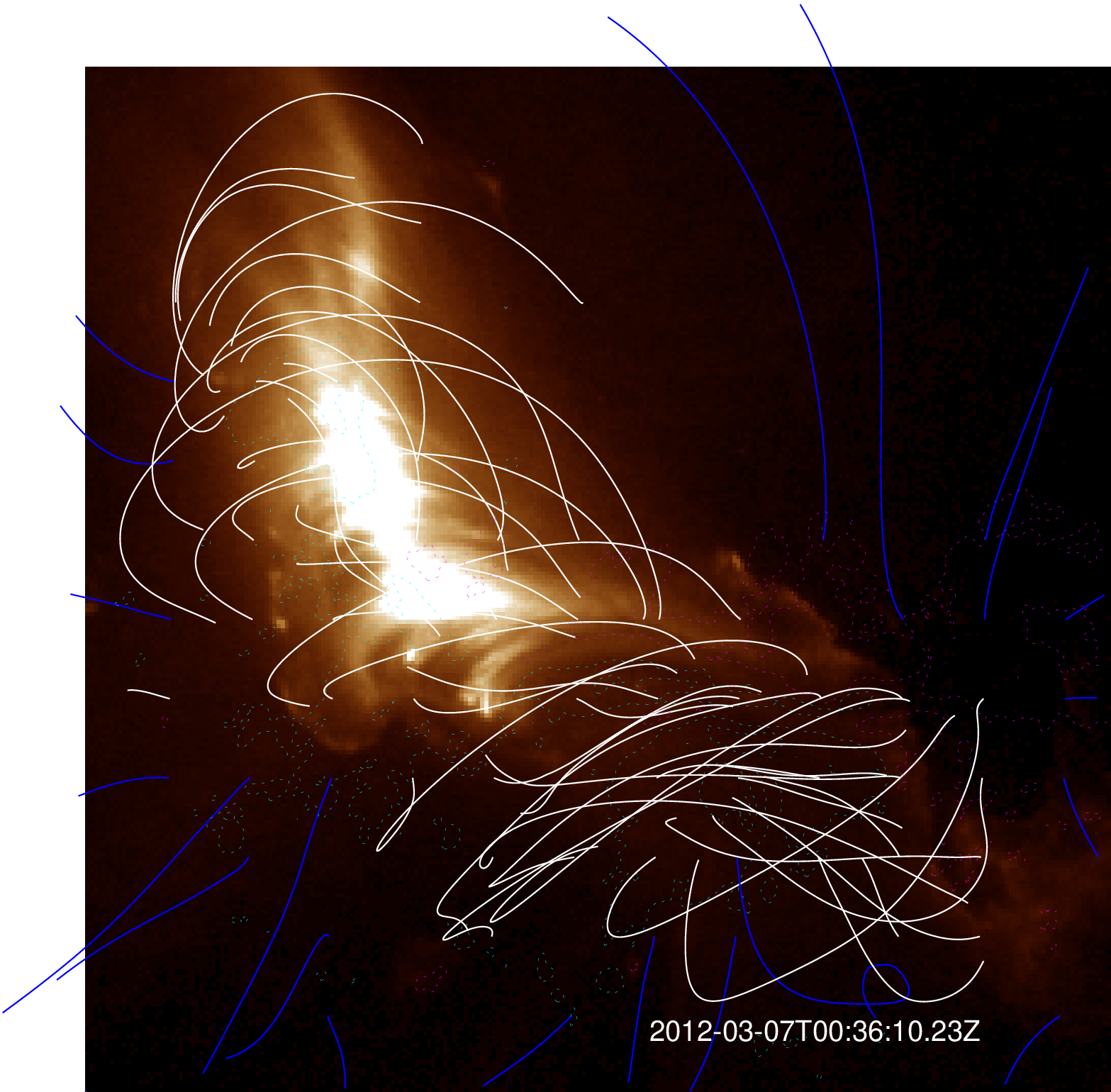}
\includegraphics[width=.32\textwidth]{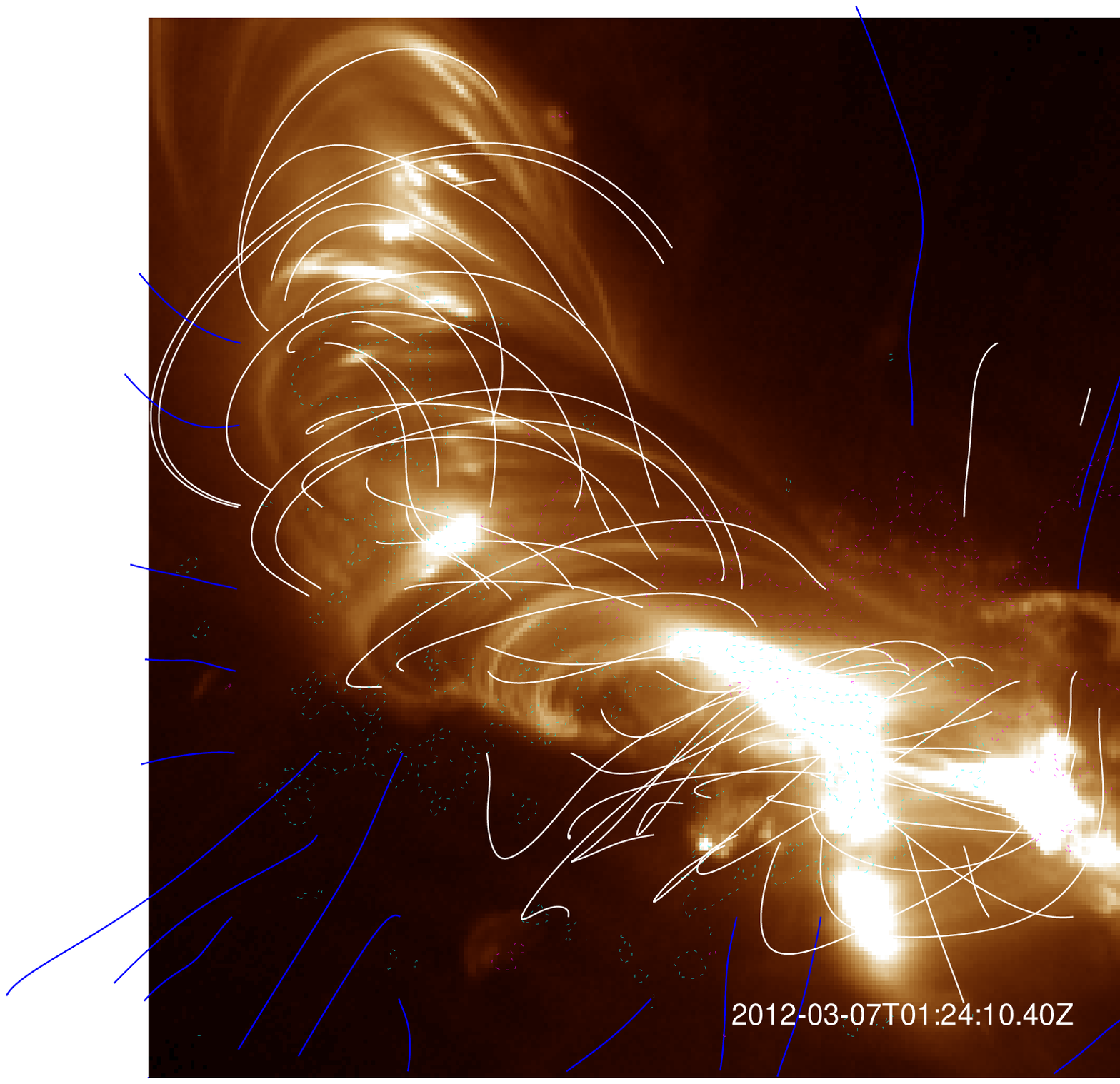}
\caption{AIA images of AR 11429 in 193 {\AA} channel observed at 00:00 UT (left), 00:36 UT (middle), and 01:24 UT (right) on March 7, which correspond to the times before the flares, post the first flare but before the second one, and post the second flare, respectively. Over-plotted are some arbitrarily chosen field lines from the NLFFF model. Contours of $\pm$500, $\pm$1300 G of ${\bf B}_z$ are also overlaid on the EUV images and are marked in cyan (negative) and pink (positive), respectively. The rough positions of the first and second eruptions, flux rope and pre-flare arcades are marked as FE1, FE2, FR and PA in the left panel, respectively. The rough positions of CS1 and CS2 are also overplotted in the left panel.\label{extrap}}
\end{figure}

A set of sample field lines from the resulting NLFFF extrapolation are displayed in Figure~\ref{extrap}, over the 193 \AA~channel images from the Atmospheric Imaging Assembly (AIA;~\citealt{2012lemen}). We have chosen three different times, namely before the first flare eruption (FE1), after FE1 but before the second flare eruption (FE2), and after FE2, respectively. The AIA images indicate that FE1 occurred in the east part of the AR and FE2 in the west part. The extrapolated field lines seem to have good alignment with the EUV background pattern. Contours of $\pm$500, $\pm$1300 G of ${\bf B}_z$ are also overlaid on the EUV images in order to identify the footpoints of the field lines. In the left panel, the brightest structure on the east side of the AR seems to correspond to a flux rope structure, and our extrapolated field lines appear to confirm this. There are some pre-flare arcades overlying the brightened structures. In the middle and right panels, a post-flare ``arcade" gradually formed and expanded.

\begin{figure}
\epsscale{.80}
\plotone{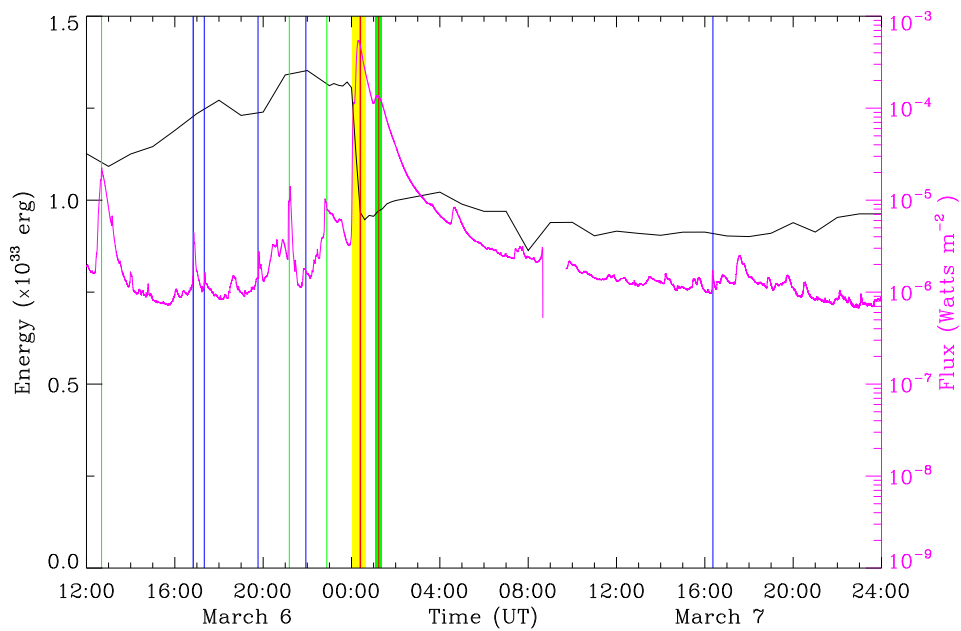}
\caption{Evolution of the magnetic free energy of AR 11429 from 12:00 UT on March 6 to 23:48 UT on March 7. The solid black line corresponds to the magnetic free energy, and the purple curve corresponds to the GOES soft-X ray flux (1-8 {\AA} channel). Vertical blue, green and red lines denote the peak times of C-, M-, and X-class flares, respectively, with their thickness roughly corresponding to the magnitude of the flare class. The vertical yellow and green squares in both panels correspond to the intervals of the first and second eruptions, respectively. \label{free_en}}
\end{figure}

Figure~\ref{free_en} shows the magnetic free energy calculated using Eq~2. The magnetic free energy was increasing before the eruptions. During the pre-flare phase from 12:00 UT on March 6 to 00:00 UT on March 7, although there were some C- and M- class flares as shown by the GOES X-ray flux, we do not see obvious decreases in the magnetic free energy. When the largest X-flare occurred at 00:02 UT on March 7, the free energy shows a dramatic decrease. The decrease in the magnetic free energy ends after the peak time of FE1, which indicates that most of the free energy was released by FE1. The amount of the energy drop during the first eruption is $~3.0\times10^{32}$ erg, accounting for about $20\%-30\%$ of the pre-flare free energy. During FE2, we observe a small increase rather than a decrease in the magnetic free energy. This is interesting as FE2 was also an X-class flare. After the two eruptions, the curve of the free energy became relatively flat.

\subsection{Evolution in Magnetic Helicity Injection} \label{mh}

There has been increasing observational evidence that the helicity of magnetic fields holds an important clue to solar flare eruptions. According to the work of Berger \& Field \citeyearpar{1984berger}, the measure of the magnetic helicity change in the solar coronal can provide physically reasonable results for torsional motions on the boundary plane. Berger \& Field \citeyearpar{1984berger} derived the Poynting theorem for the helicity in an open volume:
\begin{equation}
\frac{dH}{dt}=\oint2({\bf B}_t\cdot{\bf A}_p)v_zdS+\oint-2({\bf v}_t\cdot{\bf A}_p)B_zdS \label{dhdt},
\end{equation}
where ${\bf A}_p$ is the vector potential of the magnetic field, which is calculated by means of a fast Fourier transform method as implemented by Chae \citeyearpar{2001chae}, ${\bf v}_t$ is the tangential velocity of the real motion of the plasma on the photosphere. In Eq.~\ref{dhdt}, the first term $2({\bf B}_t\cdot{\bf A}_p)v_z$ represents the helicity injection rate via the passage of the helical field lines through the photospheric surface (i.e., emergence of new flux), and the second term $-2({\bf v}_t\cdot{\bf A}_p)B_z$ represents the helicity injection rate via the shuffling horizontal motion of the field lines on the surface (i.e., shearing or twisting motions). In the work of \citet{2001chae}, the horizontal velocity ${\bf v}_t$ is equal to the tracking velocity from the Local Correlation Tracking (LCT) method \citep{1988LCT}. However, Kusano et al. \citeyearpar{2002kusano, 2003kusano} indicated that the tracking velocity obtained from LCT is an ``image motion" of the magnetic footpoints rather than the material motion ${\bf v}_t$. Later, D\'emoulin \& Berger \citeyearpar{2003demoulin} pointed out the use of $v_z$ deduced from Doppler measurements would only duplicate part of the helicity injection rate already included in the tracking velocity. They deduced and proved that all the helicity injection rate only from the emergence of new flux can be present in the second term of Eq.~\ref{dhdt} determined by the tracking method. Here, the track velocity is determined by $-v_z{\bf B}_t/B_z$. On the other hand, the tracking method does not just measure the vertical plasma motion $v_z$, but also the horizontal motion ${\bf v}_t$. Therefore, we can present the tracking velocity ${\bf u}$ by the sum of both velocities:
\begin{equation}
{\bf u}={\bf v}_t-\frac{v_z}{B_z}{\bf B}_t.
\end{equation}
Namely, the tracking method provides the transverse velocities including both of the effects of the shearing motion and the vertical motion. This method can only compute the total helicity injection rate across the photosphere as the equation below \citep{2003demoulin}:
\begin{equation}
\frac{dH}{dt}=-2\oint({\bf u}\cdot{\bf A}_p)B_zdS \label{dhdt2}.
\end{equation}
We determine the tracking velocity ${\bf u}$ of the magnetic fields at their photospheric footpoints using a Fourier Local Correlation Tracking (FLCT) method \citep{2008ASPC..383..373F}, which is the upgraded version of LCT. When we use FLCT to measure the tracking velocity, some appropriate parameter settings will make the results more accurate. In order to calculate the tracking velocity, we need to give the time interval $\Delta t$ between two magnetograms for FLCT. Here it is 720 s. We also use a Gaussian windowing function as weighting function for sub-images in the tracking method. The parameter $\sigma$ as the width of Gaussian function is 15, which has been proved to be the best setting in the numerical experiment of Welsch et al. \citeyearpar{2007welsch}.

Figure~5 shows the helicity injection rate as a function of time, which is determined by Eq.~\ref{dhdt2}. The rate is determined every 12 minutes for 36 hrs. The rate shows a considerable fluctuation. Before FE1, the AR accumulated the negative helicity. Then the negative injection rate decreased around FE1 and changed its sign. Until FE2 the helicity injection rate reached to its positive peak, about $-4.0\times10^{41}$ Mx$^2$ hr$^{-1}$. After FE2, the helicity injection decreased and changed its sign again to negative and kept this balanced stage with a considerable fluctuation. It suggests some possibility that the changes of the sign of the helicity injection rate might be associated with the eruptions. Similar results can be found in the work of \citet{2003kusano}.

Figure~\ref{helicity} shows the accumulated change of the magnetic helicity, which was obtained by integrating the measured $dH/dt$ from the start of the observing run to the specified time. The black curve represents the accumulated change of the magnetic helicity from the helicity injection rate of Eq.~\ref{dhdt2}. The negative accumulated change of the injection rate kept an approximately constant increase until FE1, then because of the change of the sign of the injection rate (see Figure~5), it stopped increasing. After FE2, it increased again in a relative slower growth rate. The increase of the magnetic helicity should be attributed to the effects of both the emergence and the shearing motions.

\section{Coronal Magnetic Field Restructuring due to Implosion} \label{discussion}

A prominent feature during the eruptions is the magnetic $implosion$ phenomenon. This scenario was proposed by Hudson \citeyearpar{2000hudson} for the first time. The energy for coronal transient phenomena comes from the stressed coronal magnetic field. According to Hudson \citeyearpar{2000hudson}, the energy release corresponds to a reduction in the magnetic pressure, and the unbalanced forces in the magnetic field will cause the magnetic structure to contract, i.e., the appearance of the implosion phenomenon.

Figure~7 shows the evolution of the NLFFF extrapolated magnetic field component perpendicular to the slice as marked in Figure~1. An obvious contraction of the magnetic fields before FE1 can be seen in the upper panels corresponding to the slice CS1. This is consistent with the change in the photospheric horizontal field shown in Figure~\ref{magfluxPIL}. Note that the times selected for CS1 are before the peak time of FE1. The contraction of the magnetic field may be caused by the relaxation of the twisted magnetic field, and is likely part of the energy release process during FE1. The evolution of the magnetic fields is consistent with the conjecture of Hudson \citeyearpar{2000hudson}. The lower panels of Figure~7, however, indicate that magnetic contraction did not occur during FE2.

In order to find out more details about the magnetic field change during the two flares, we calculate the magnetic flux for the ten cross sections (or slices) of the extrapolated coronal magnetic field data cube (as shown by the white, red and blue lines in Figure~\ref{magmap}). The magnetic flux $F_\perp$ perpendicular to the cross sections and the flux $F_\parallel$ parallel to the cross sections are shown in Figure~\ref{flux_x} and Figure~\ref{flux_y}, respectively. Here we define the flux $F_\parallel$ as the integration of the absolute value of the parallel magnetic field, which consists of the north-south and vertical components of the magnetic field vector, over the cross section. Looking at all plots of these slices in Figure~8 and 9, the most prominent changes of the magnetic field, all happened in plots (c), (d), and (e), which should correspond to the source region of FE1 (see Figure~1) and show the probable position and range of the source region. During FE1, the decrease in $F_\perp$ (see Figure~8) in this region may be owing to the ejection of the flux rope structure associated with the first CME. This process is so much like the ``tether-cutting'' model \citep{2001moore}, i.e., the flux rope expanded upward and two-ribbon flare near the solar surface formed at the same time. This may interpret why $F_\perp$ decreased and $F_\parallel$ increased (see Figure~9) in plots (c), (d), and (e). On the other hand, the magnetic flux through each slice increased during FE1. Since the computation region is in a relatively low altitude, we suggest that it reflects the magnetic contraction during FE1, i.e., the field lines decrease in length from the higher to lower altitudes, which increases the magnetic field density in the lower corona as suggested by Hudson \citeyearpar{2000hudson}. This can be visually seen in the upper panels of Figure~\ref{contraction}. Compared with the field changes during FE1, the magnetic flux changes during FE2 were small (see Figures~8 and 9). Specifically, no $implosion$ phenomenon occurred during FE2. This is consistent with the lower panels of Figure~7.

From Figure~8 and 9, the difference of the variations of the magnetic field between the eruptions with implosion (FE1) and without implosion (FE2) can be well observed. During FE1, generally speaking, the energy was released (see Figure~4) for radiation or other observation phenomena, the magnetic flux should decrease. However, contrary to what we think, the total magnetic flux near the source region increased (see Figure~9). This is interesting if we use the implosion theory to explain it, i.e., if part of the coronal field expand to higher corona, a further compensating implosion in the lower corona will simultaneously take place, and the magnetic pressure inward may make the lower magnetic field become more compact and the magnetic flux in lower corona increase (seen as (c), (d), and (e) in Figure~9). By contrast, there was no implosion phenomenon during FE2 and the magnetic flux changed as usual.
 \section{Conclusions} \label{conclusion}

We have analyzed in detail 36 hours of 12-minute SDO/HMI vector magnetic field observations covering the X5.4 and X1.3 successive flares at 00:24 UT and 01:14 UT on 2012 March 7, respectively. By means of an advanced NLFFF extrapolation method, we derived the coronal magnetic fields and magnetic free energy from the preprocessed boundary conditions. The magnetic helicity was computed using the HMI vector magnetograms. Through the extrapolated coronal magnetic field, magnetic free energy and helicity, we investigated the magnetic field restructuring associated with the two prominent successive eruptions. The main conclusions are:

1. Near the PIL region, the photospheric vector fields became more horizontal after the first flare than the preflare state. It shows a stepwise increase in the photospheric horizontal field component.

2. The magnetic free energy shows a stepwise decrease during the first flare while no apparent change during the second one. The amount of the energy drop during the first eruption is $~3.0\times10^{32}$ erg, accounting for about $20\%-30\%$ of the pre-flare free energy.

3. The helicity injection rate changed the sign from negative to positive, reaching its positive peak about $-4.0\times10^{41}$ Mx$^2$ hr$^{-1}$ during the eruptions. It suggests that the changes of the sign of the helicity injection rate might be associated with the eruptions.

4. The extrapolated coronal magnetic field shows a contraction behavior during the first eruption. This is consistent with the $implosion$ process suggested by Hudson \citeyearpar{2000hudson}. Meanwhile, the magnetic field changes of the successive eruptions with implosion and without implosion were well observed in the same AR.

\acknowledgments

We are grateful to Dr. Thomas Wiegelmann for his generous sharing of his latest version of the NLFFF optimization code. The work was supported by the Specialized Research Fund for State Key Laboratories of China, the Recruitment Program of Global Experts of China under grant Y3B0Z1A840, the Strategic Priority Research Program on Space Science from the Chinese Academy of Sciences (XDA04060801), and the CAS Key Research Program KZZD-EW-01. We acknowledge the use of data from SDO.

\clearpage

%% the file, uncomment the \plotone, \plottwo, and
%% If you need a layout that cannot be achieved with \plotone or
%% \plottwo, you can invoke the graphicx package directly with the
%% \includegraphics command or use \plotfiddle. For more information,
%% please see the tutorial on "Using Electronic Art with AASTeX" in the
%% documentation section at the AASTeX Web site, http://aastex.aas.org/
%%
%% The examples below also include sample markup for submission of
%% supplemental electronic materials. As always, be sure to check
%% the instructions to authors for the journal you are submitting to
%% for specific submissions guidelines as they vary from

%% This example uses \plotone to include an EPS file scaled to
%% 80% of its natural size with \epsscale. Its caption
%% has been written to indicate that additional figure parts will be
%% available in the electronic journal.

\begin{figure}
\epsscale{.80}
\plotone{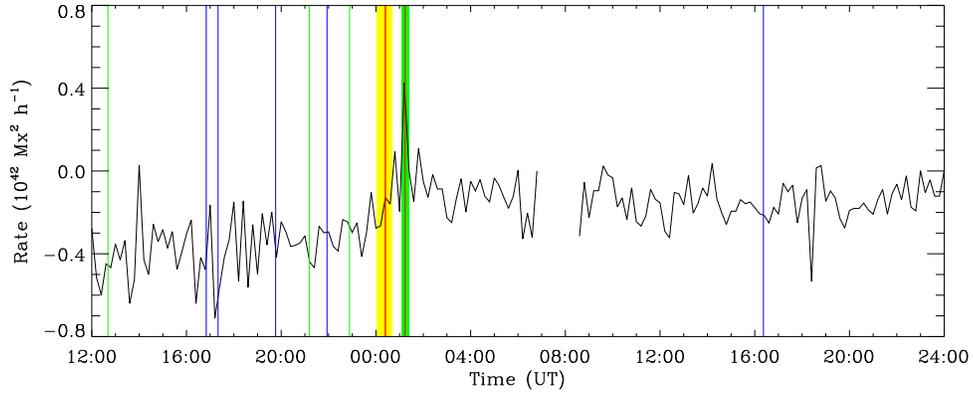}
\caption{Helicity injection rate as a function of time, which is determined by Eq.~\ref{dhdt2}. The vertical lines and squares have the same meaning as in Figure~4. \label{helicity_rate}}
\end{figure}

\begin{figure}
\epsscale{.80}
\plotone{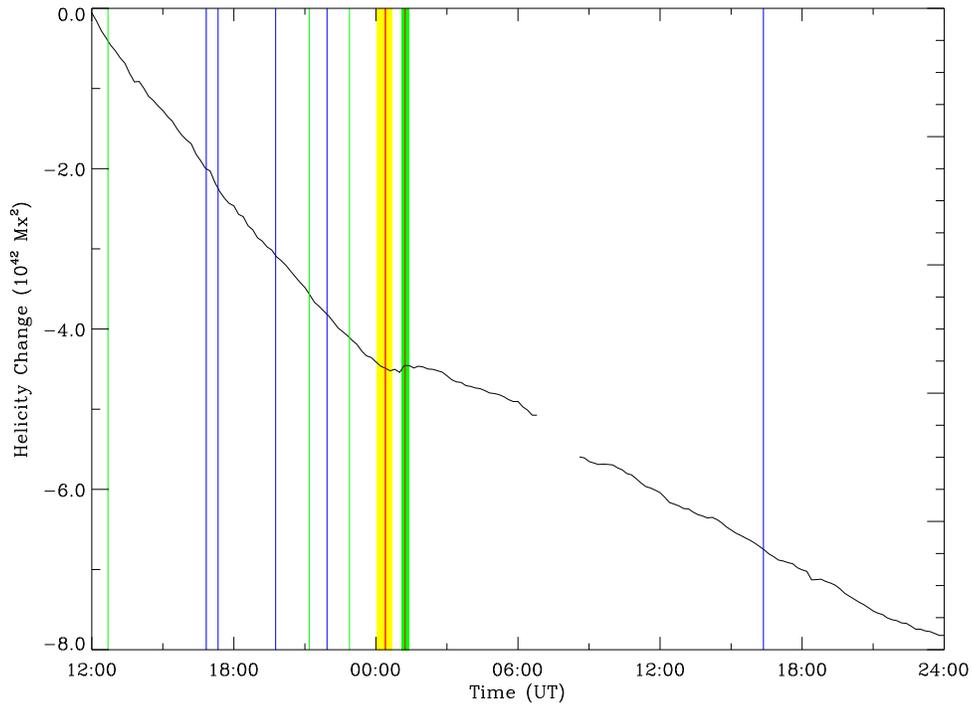}
\caption{Accumulated change of the magnetic helicity as a function of time. The vertical lines and squares have the same meaning as in Figure~4.\label{helicity}}
\end{figure}

\begin{figure}
\epsscale{.80}
\plotone{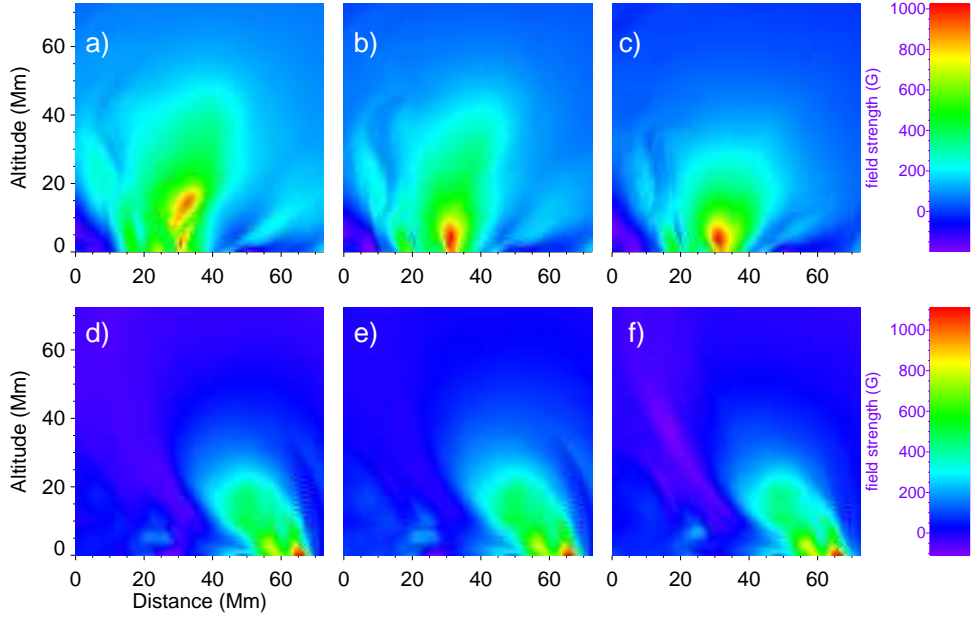}
\caption{Distributions of the horizontal NLFFF field ($B_h$) in two vertical cross sections (CS1 and CS2 as shown in Figure~1). Only the component perpendicular to the cross section is shown. a, b, c: Distribution of the horizontal field component in CS1 at 23:48 UT on March 6, 00:12 UT and 00:24 UT on March 7, respectively. d, e, f: Distribution of the horizontal field component in CS2 at 01:00 UT, 01:24 UT, and 02:00 UT on March 7, respectively. \label{contraction}}
\end{figure}

\begin{figure}
\epsscale{.80}
\plotone{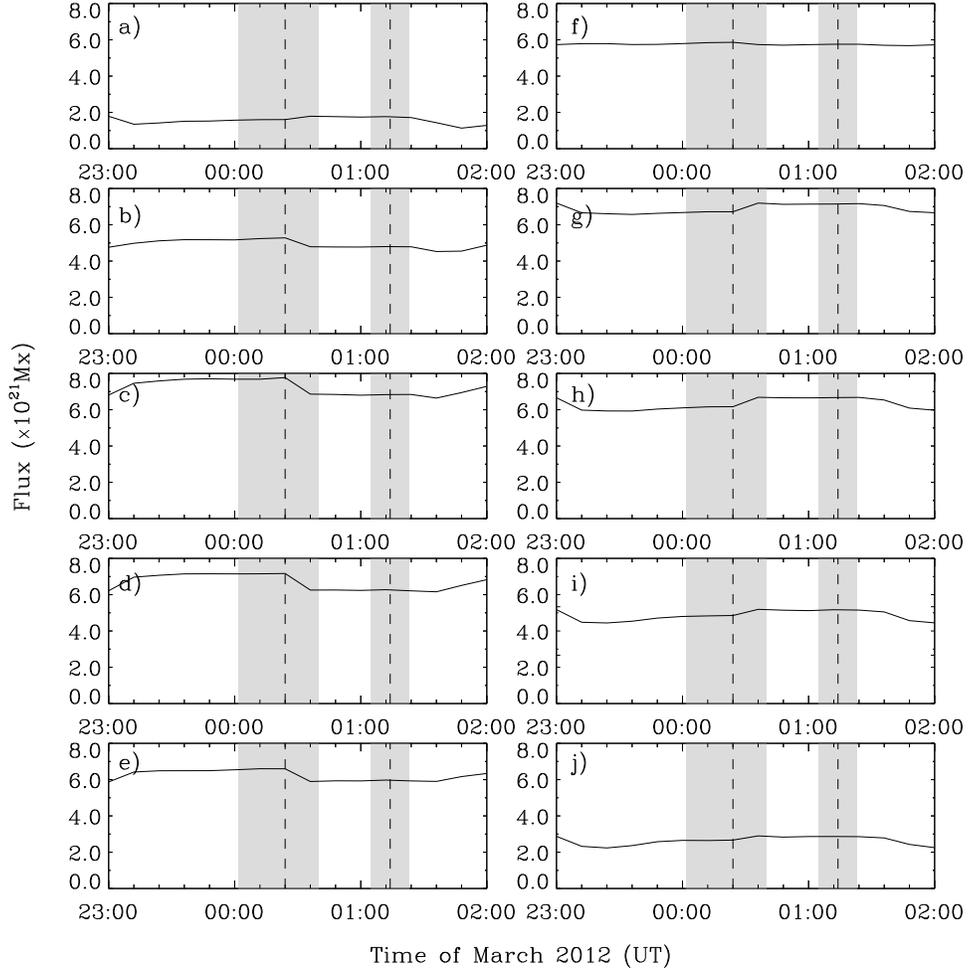}
\caption{The magnetic flux of the field components perpendicular to the cross sections from 23:00 UT on March 6 to 02:00 UT on March 7. The ten panels from (a) to (j) correspond to the ten cross sections from left to right in Figure~\ref{magmap}, respectively. The two gray squares indicate the times of the first and second eruptions from their beginning to end, and the vertical lines mark the peak times of FE1 and FE2. \label{flux_x}}
\end{figure}

\begin{figure}
\plotone{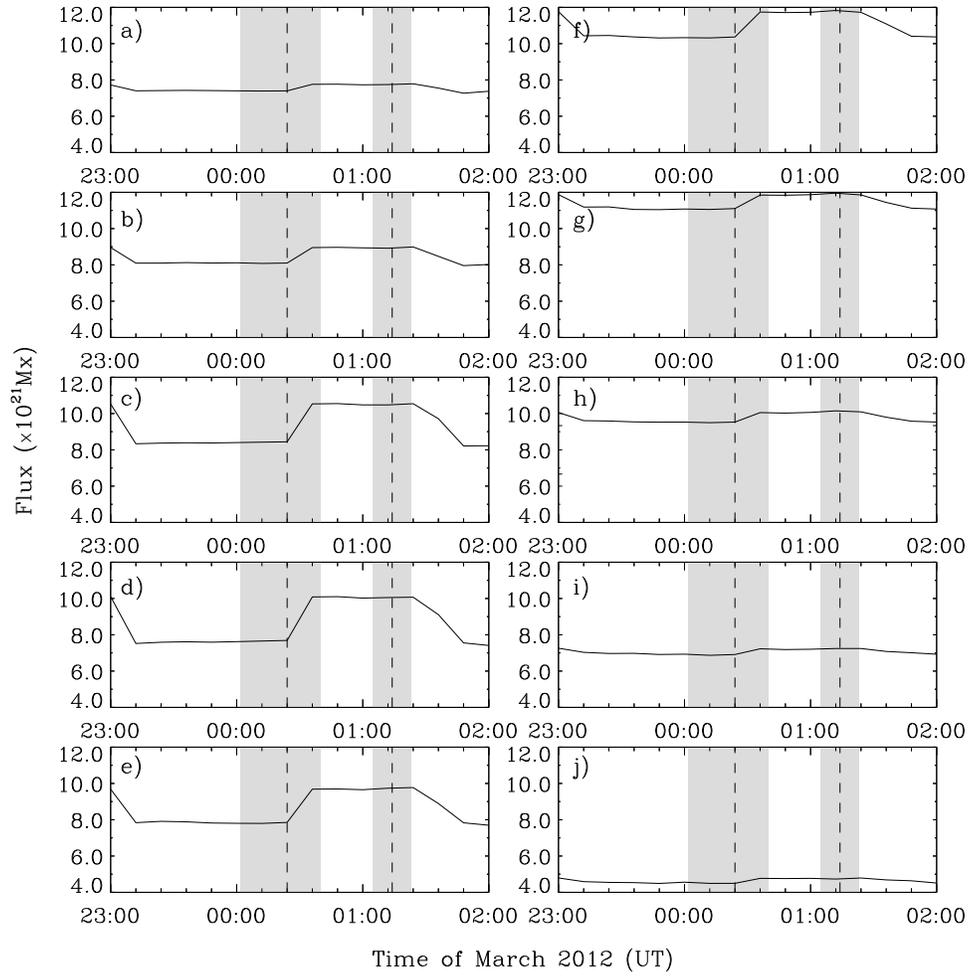}
\caption{Similar to Figure~\ref{flux_x}, but for the magnetic flux of the field components parallel to the cross sections.\label{flux_y}}
\end{figure}

\clearpage

%% Tables may also be prepared as separate files. See the accompanying
%% sample file table.tex for an example of an external table file.
%% To include an external file in your main document, use the \input
%% command. Uncomment the line below to include table.tex in this
%% sample file. (Note that you will need to comment out the \documentclass,
%% \begin{document}, and \end{document} commands from table.tex if you want
%% to include it in this document.)


%% The following command ends your manuscript. LaTeX will ignore any text

\end{document}